\documentclass[fdp,a4paper,fleqn%
]{w-art}
\usepackage{times,cite,w-thm}
\usepackage[]{graphicx}
\usepackage[latin1]{inputenc}
\usepackage[breaklinks=true,colorlinks=false,linkcolor=black,urlcolor=black,pagecolor=black]{hyperref}
\usepackage{amsfonts,amssymb,amsmath,amsthm,bbm}
\usepackage{xcolor}
\usepackage{graphicx}
\newcommand{\red}[1]{{\color{red} #1 \color{black}}}
\newcommand{\blue}[1]{{\color{blue} #1 \color{black}}}

\def\be{\begin{equation}}
\def\ee{\end{equation}}
\def\bea{\begin{eqnarray}}
\def\eea{\end{eqnarray}}
\def\beq{\begin{equation}}
\def\eeq{\end{equation}}
\def\beqa{\begin{eqnarray}}
\def\eeqa{\end{eqnarray}}

\def\t{\tau}

\begin{document}
\DOIsuffix{theDOIsuffix}
\Volume{}
\Month{}
\Year{}
\pagespan{1}{}
\Receiveddate{XXXX}
\Reviseddate{XXXX}
\Accepteddate{XXXX}
\Dateposted{XXXX}



\title[Holographic Cosmological Deconfinement]{Holographic Cosmological Backgrounds, Wilson Loop (De)con\-finement and Dilaton Singularities}


\author[J.~Erdmenger]{Johanna Erdmenger\inst{1,}%
\footnote{E-mail:~\textsf{jke@mppmu.mpg.de}  
            }}
\address[\inst{1}]{Max-Planck-Institut f\"ur Physik, 
F\"ohringer Ring 6, 80805 M\"unchen, Germany}
\author[K.~Ghoroku]{Kazuo Ghoroku\inst{2,}%
\footnote{E-mail:~\textsf{gouroku@fit.ac.jp}  
            }}
\address[\inst{2}]{Fukuoka Institute of Technology, Wajiro, Higashi-ku
Fukuoka 811-0295, Japan}
\author[R.~Meyer]{Ren\'e Meyer\inst{3,}%
\footnote{Corresponding author and conference speaker, \quad E-mail:~\textsf{meyer@physics.uoc.gr},
            Phone: +30\,2810\,39\,4233, 
            Fax: +30\,2810\,39\,4274. The work of RM was in part supported by the European grants
FP7-REGPOT-2008-1: CreteHEPCosmo-228644 and  PERG07-GA-2010-268246, as well as the EU Program ``Thalis" ESF/NSRF 2007-2013. RM gratefully acknowledges the support and hospitality of the Korean Institute for Advanced Study (KIAS), of the Center for Quantum Spacetime at Sogang University (Seoul) and of the Isaac Newton Institute for Mathematical Sciences during the final stages of this work. IP wishes to thank the University of Crete for the hospitality, where part of this work was carried out.}}
\address[\inst{3}]{Physics Department, 
University of Crete, 
71003 Heraklion, Greece and\\
Korea Institute for Advanced Study, Seoul 130-012, Korea}
\author[I.~Papadimitriou]{Ioannis Papadimitriou\inst{4,}%
\footnote{E-mail:~\textsf{Ioannis.Papadimitriou@csic.es}\hfill CCTP-2012-04, IFT-UAM/CSIC-12-26, KIAS-P12021
            }}
\address[\inst{4}]{Instituto de F\'{\i}sica Te\'orica UAM/CSIC, Universidad Aut\'onoma de Madrid, Madrid 28049, Spain.}
\begin{abstract}
We review a construction of holographic geometries dual to
$\mathcal{N}=4$ SYM theory on a Friedmann-Robertson-Walker background
and in the presence or absence of a gluon condensate and instanton
density. We find the most general solution with arbitrary scale factor
and show that it is diffeomorphic to topological black holes. We
introduce a time-dependent boundary cosmological constant $\lambda(t)$ and show energy-momentum conservation in this background. For constant $\lambda$, the deconfinement properties of the temporal Wilson loop are analysed. In most cases the Wilson loop confines throughout cosmological evolution. However, there is an exceptional case which shows a transition from deconfinement at early times to confinement at late times. We classify the presence or absence of horizons, with important implications for the Wilson loop.
\end{abstract}
\maketitle                   





%

\section{Background and Holographic Stress Energy Tensor}

In AdS/CFT, the holographically defined theory is only sensitive to the conformal class of the boundary metric, not the metric itself.\footnote{In the presence of a conformal anomaly observables depend on the chosen representative of the conformal class, but transform in a well-defined way, governed by the anomaly, into each other \cite{IPThermo}.} It is hence possible to put any holographically defined flat space field theory on  $dS_4$ or $AdS_4$ via an appropriate boundary conformal transformation, amounting to particular bulk diffeomorphisms. The same holds for any metric of Friedmann-Robertson-Walker (FRW) type,
\be\label{FRW}
ds_{(0)}^2 = - dt^2 + a_0^2(t) d\Omega_k^2\,.
\ee
It is hence conceivable that holographic duals for field theories
living on an FRW universe can be constructed in such a way. Such a
construction from 5D Einstein-Hilbert theory has been put forward in
\cite{Tetradis}, describing a universal sector of any holographic
theory with a stress-energy tensor.\footnote{For more details on the
  effective holographic approach to model building, cf.~\cite{CGKKM}.}
In the following we review such a construction \cite{EGM} based on the
background of \cite{LT}. This $\mathcal{N}=2$ supersymmetric
background describes $\mathcal{N}=4$ SYM theory in a vacuum of
constant (anti) self-dual instanton density, and is known to lead to
Wilson loop confinement. We will first review the derivation of the
cosmological background presented in \cite{EGM}, and then show its
diffeomorphism equivalence to the static geometry of \cite{LT}. {In
  this course we will find a more general class of solutions including
  the one of \cite{EGM}, clarify issues regarding the implications of
holographic stress energy tensor conservation, comment on the meaning of $a_0(t)$ and the boundary cosmological constant $\lambda(t)$ from a bulk perspective, and classify the horizons appearing in the cosmological solution. In sec.~\ref{qqbar} we then classify the (de)confinement properties of the Wilson loop in this background, and compare to the AdS-Schwarzschild case.}

Ten-dimensional type IIB supergravity truncates to the metric $g$, dilaton $\Phi$, axion $\chi$ and five form $F_{(5)}$,
\beq
 S={1\over 2\kappa^2}\int d^{10}x\sqrt{-g}\left(R-
{1\over 2}(\partial \Phi)^2+{1\over 2}e^{2\Phi}(\partial \chi)^2
-{1\over 4\cdot 5!}F_{(5)}^2
\right)\,. \label{10d-action}
\eeq
The other IIB fields are consistently set to zero, and  
$\chi$ is Wick rotated \cite{GGP}. Under the Freund-Rubin ansatz   
$F_{\mu_1\cdots\mu_5}=-\sqrt{\Lambda}/2~\epsilon_{\mu_1\cdots\mu_5}$  
\cite{KS2,LT} and  $\mathcal{M}_{10} = \mathcal{M}_5\times S^5$, 
 the action \label{10d-action} reduces to 
\beq
 S={1\over 2\kappa_5^2}\int d^5x\sqrt{-g}\left(R+3\Lambda-
{1\over 2}(\partial \Phi)^2+{1\over 2}e^{2\Phi}(\partial \chi)^2
\right)\,. \label{5d-action}
\eeq
Its equations of motion have an $\mathcal{N}=2$ supersymmetric solution if \cite{GGP,KS2} 
\beq\label{ansa1}
     \chi=-e^{-\Phi}+\chi_0\ ,
\eeq
for some constant $\chi_0$. 
Under this Ansatz the metric decouples from the axion-dilaton,
\bea\label{gravity}
 R_{MN}&=&-\Lambda g_{MN}\,,\quad 0 = \partial_M\left(\sqrt{-g}g^{MN}\partial_N e^{\Phi}\right)\,.
\eea
In \cite{EGM} we followed \cite{Binetruy} and examined time-dependent solutions of the form
\beq
ds^2_{\rm E}=-n^2(t,y)dt^2+a(t,y)^2d \Omega_k^2+dy^2\,,\quad i,j=1,\dots,3. 
\label{y-co}
\eeq
The $tt$ and $yy$ components of \eqref{gravity} become \cite{Binetruy} 
\beq\label{cosmo}
 \left({\dot{a}\over na}\right)^2+{k\over a^2}=
   -{\Lambda\over 4}+\left({a'\over a}\right)^2
  +{C\over a^4}\,,\quad \dot{a}=\partial a/\partial t\,,\quad
  a'=\partial a/\partial y \, .
\eeq
This first order equation comes from integrating Einstein's equations (see \cite{Binetruy,Lang} for details). 
The \textbf{\textit{dark radiation constant}} $C$ appears in \eqref{cosmo} as a  radiation term ${C\over a^4}$ \cite{Lang,Kehagias:1999vr}. The  remaining $ty$ component reads
\beq\label{tyequation}
0 = \frac{n'}{n}\frac{\dot a}{a} - \frac{\dot a'}{a} \quad\Rightarrow \quad  n(t,y)={\dot{a}(t,y)\over \eta(t)}\,.
\eeq
\textbf{General Solution and Time Independence of $\lambda$}
The most general solution of (\ref{cosmo}) and (\ref{tyequation}) depends onn two arbitrary functions of time, $\theta(t)$ and $\eta(t)$, 
\be\label{gensol}
a^2=\frac{C+\Lambda^{-1}\left(\Lambda\theta(t)e^{\sqrt{\Lambda}y}-k-\eta^2(t)\right)^2}{\Lambda\theta(t)e^{\sqrt{\Lambda}y}},\quad n=\frac{\dot a}{\eta(t)}\,.
\ee
With $\sqrt{\Lambda}=2/R$, $\exp(\sqrt{\Lambda} y)=(r/R)^2$, $\theta(t)=a_0 ^2(t)$, and $\eta(t)=\dot a_0(t)$, the solution for $a^2 = a_0(t)^2 A(t,y)^2$ is identical to eq.~(31) of \cite{EGM}. This choice of $\theta$ and $\eta$ also fixes the boundary metric to be \eqref{FRW}. Eq.~(\ref{cosmo}) then becomes
\beq\label{A}
 \left({\dot{a}_0\over a_0}\right)^2+{k\over a_0^2}=
   -{\Lambda\over 4}A^2+\left({A'}\right)^2
  +{C\over a_0^4 A^2}\,,\quad A'=\partial A/\partial y\,.
\eeq
The \textbf{main observation of \cite{EGM}} is that the LHS of \eqref{A} is a function of $t$ only, depending on the integration function $a_0(t)$. Eq.~\eqref{A}  thus can be solved by introducing a (\textit{a priori} time-dependent) \textbf{\textit{boundary cosmological constant}} $\lambda(t)$,
\beq
 \left({\dot{a}_0\over a_0}\right)^2+{k\over a_0^2} = \lambda(t)\,,\quad \lambda(t) =  -{\Lambda\over 4}A^2+\left({A'}\right)^2
  +{C\over a_0^4 A^2}\,.
\eeq
The LHS of \eqref{A} becomes Friedmann's equation,  while the RHS can be integrated to yield $A$ as in eq.~(31) of \cite{EGM}, with $n$ given by \eqref{tyequation}. {Note that in this treatment the boundary cosmological constant $\lambda(t)$, and hence the cosmological evolution $a_0(t)$, is essentially \textit{arbitrary}: From the point of view of the bulk $a_0(t)$ is, upon requiring an FRW boundary metric, a single free integration function which is not determined in the model itself. Dynamical determination of $\lambda(t)$ by coupling to boundary gravity or some other mechanism may be an interesting direction of future research. In the remainder of this paper we however \textit{choose $\lambda$ to be time-independent} in order to investigate the properties of the dual gauge theory in the usual cosmological backgrounds (dS/flat/AdS). It should also be noted that only for a time-independent $\lambda$ the solution for $n(t,y)$ in eq.~(33) of \cite{EGM} coincides with \eqref{gensol}.}\\

\textbf{Holographic Stress-Energy-Tensor} In the above construction, the time-dependence of $\lambda(t)$ was \textit{a priori} unconstrained. Using the results of App.~B of \cite{AxioDilaton}, the vacuum expectation value of the boundary stress-energy tensor can be found for a general $\lambda(t)$, but its form is complicated and not very illuminating. For a time-independent cosmological constant $\lambda$ it greatly simplifies to the following ideal fluid form
\beq\label{Tmn}
\hspace{-0.85cm}
\langle T^\mu{}_\nu \rangle=\text{diag}(-\rho,p,p,p),\  
\rho = \frac{12R^3}{16\pi G_N^{(5)}} \left( \frac{C }{4R^2 a_0^4(t)} + \frac{\lambda^2}{16} \right),\ p = \frac{4R^3}{16\pi G_N^{(5)}} \left( \frac{C }{4R^2 a_0^4(t)} - \frac{3\lambda^2}{16}\right).
\eeq
The form of \eqref{Tmn} is interesting by itself: The part
proportional to the dark radiation constant $C$ is actually conformal,
and dilutes with $a_0^{-4}$, as relativistic radiation in an expanding
FRW universe does. {We will see below that this is not a coincidence: In
  the frame where \eqref{y-co} becomes the usual static topological
  black hole, $C$ is the black hole mass parameter, and contributes to
  \eqref{Tmn} relativistic $\mathcal{N}=4$ radiation in an expanding
  universe.} The cosmological constant part on the other hand is not
conformal, and originates from the conformal anomaly of the
$\mathcal{N}=4$ SYM fields on the curved FRW space-time
\cite{EGM}. {We note that the holographic
  stress energy tensor as e.g. calculated from eq.~(B19) of
  \cite{AxioDilaton} is covariantly conserved by construction for
  arbitrary time-dependent $\lambda(t)$. Stress energy conservation
  hence does not restrict $\lambda(t)$ to be time-independent,
 as originally claimed in \cite{EGM}. 

\textbf{Equivalence with Static Topological Black Holes} We now show that the solutions obtained above are in fact diffeomorphism equivalent to the usual static form of topological black holes in $AdS_5$. Having explicitly constructed the function $a(y,t)$, we now consider a diffeomorphism of the form 
$(y,t)\mapsto(a(y,t),\t(y,t))$, 
with $\t(y,t)$ to be specified. Implementing such a diffeomorphism the metric 
(\ref{y-co}) becomes 
\be
ds^2=\frac{1}{\left(a'\frac{\dot\t}{n}-\t'\frac{\dot a}{n}\right)^2}
\left[\left(\frac{\dot\t^2}{n^2}-\t'^2\right)da^2-P(a)d\t^2
+2\left(a'\t'-\frac{\dot a\dot\t}{n^2}\right)dad\t\right]+a^2d\Omega_k^2\,,
\ee
where $P(a)\equiv k + \frac{\Lambda}{4} a^2 - \frac{C}{a^2}$. Absence of a mixed term requires that there
exists a function $f(a)$ such that  
\be\label{mixed}
\t'=\frac{\dot a}{n}f(a),\quad \frac{\dot\t}{n}=a'f(a), 
\ee
The function $\t(y,t)$ therefore exists iff there exists a function $f(a)$ that satisfies the integrability condition 
\be
\label{integrability-condition}
\partial_t\left(\frac{\dot a}{n}f(a)\right)=\partial_y\left(na'f(a)\right).
\ee
Using \eqref{cosmo} and \eqref{tyequation} we find that this condition is satisfied for 
$f(a)=1/P(a)$. From \eqref{mixed} it follows that 
\begin{eqnarray}\nonumber
\sqrt{\Lambda}\tau(y,t)&=&\frac{1}{\sqrt{k^2+\Lambda C}}
\left[\xi_-{\rm log}\left(\frac{\Lambda\theta(t)e^{\sqrt{\Lambda}y}-\xi_+^2-(\eta(t)+\xi_-)^2}
{\Lambda\theta(t)e^{\sqrt{\Lambda}y}-\xi_+^2-(\eta(t)-\xi_-)^2}\right)\right.\\\label{tau}
&&\phantom{}\left.-i\xi_+{\rm log}\left(\frac{\Lambda\theta(t)e^{\sqrt{\Lambda}y}+\xi_-^2-(\eta(t)+i\xi_+)^2}
{\Lambda\theta(t)e^{\sqrt{\Lambda}y}+\xi_-^2-(\eta(t)-i\xi_+)^2}\right)\right]
+\int^t\frac{d\bar t\dot{\theta}(\bar t)}{\eta(\bar t)\theta(\bar t)},
\end{eqnarray} 
where $\xi_\pm=\sqrt{\left(\sqrt{k^2+\Lambda C}\pm k\right)/2}$. In the $(a,\tau)$ frame 
the metric is the \textbf{static topological black hole}
\be\label{static-metric}
ds^2= \frac{da^2}{k + \frac{\Lambda}{4} a^2 - \frac{C}{a^2}}-\left( k + \frac{\Lambda}{4} a^2 - \frac{C}{a^2}\right)d\t^2+a^2d\Omega_k^2.
\ee
Although the existence of such a diffeomorphism has been discussed before \cite{Tetradis}, to our knowledge this is the first time that it is constructed explicitly. In particular, this shows that the dark radiation constant is the black hole mass parameter. This is in accordance with the form of the stress energy tensor \eqref{Tmn}, where the dark radiation contribution is conformal and dilutes in the same way as relativistic radiation ($\propto a_0^{-4}$). \\

\textbf{The Dilaton Solution} Assuming spatial homogeneity the dilaton equation 
in the $(a,\tau)$ frame reads
\be
\label{dilaton-eq}
a^{-3}P(a)\partial_a\left(a^3P(a)\partial_a e^\Phi\right)-\partial_\t^2e^\Phi=0.
\ee
The general static solution in these coordinates is
\bea\label{dilsol}
e^{\Phi}
&=&e^{\phi_0}+q
\left(\xi_+^2\log\left(1-\frac{4\xi_-^2}{\Lambda a^2}\right)+\xi_-^2\log\left(1+\frac{4\xi_+^2}{\Lambda a^2}\right)\right),
\eea
where $\phi_0$ and $q$ are arbitrary constants, related to the asymptotic value of the dilaton $g_s$ and the asymptotic D(-1) charge $Q$ (which has to be positive for positive $e^\Phi$) via
\be\label{asymptotics}
e^\Phi \simeq g_s + \frac{Q(t)}{r^4} + {\cal O}(r^{-6})\,,\quad g_s=e^{\phi_0}  \,,\quad 
Q(t)=-\frac{qC^2R^4\sqrt{k^2+\Lambda C}}{2a_0^4}.
\ee
Some special time dependence can be taken into account 
by allowing $e^{\phi_0}$ and $q$ to be linear functions of $\tau$. The general time-dependent solution 
of (\ref{dilaton-eq}) can be constructed by Fourier transforming the dilaton in $\tau$ and solving 
numerically the resulting second order linear ODE. 

A few comments are in order here. First, the solution obtained above in the $(a,\tau)$ coordinate system contains very non-trivial time dependence when transformed back into the $(y,t)$ coordinates. This 
is to be contrasted with the approximate expression given in eq.~(41) of \cite{EGM}, where it is assumed that 
$a_0(t)$ varies very slowly in time so that it can be taken to be a constant. Of course, this approximation
becomes exact in the case $k=0$ and $\lambda=0$. Secondly, unless $q=0$, the dilaton diverges logarithmically at the static black hole horizon corresponding to the real root of $P(a)=0$, namely $\Lambda a^2=4\xi_-^2$, which is nothing but the dilaton singularity of Liu and Tseytlin \cite{LT}. However, as we will discuss below, the horizons in the $(a,\tau)$ and $(y,t)$ coordinate systems do {\em not} coincide. We will discuss the implications of the dilaton singularity after an analysis of the horizons below.\\

\textbf{Classification of Horizons} In both $(y,t)$ and $(a,\tau)$ coordinate systems the horizons are time dependent curves, but in fact, these curves do not coincide in the two coordinate systems. The (naive) horizons in the $(y,t)$ coordinates are defined by the condition $n(y,t)=0$, which yields
\be
\left(\frac{r_\pm}{R}\right)^2=\frac{\lambda R^2}{4}\pm\frac{R\sqrt{C}}{2a_0^2(t)}.
\ee 
However, the horizons of the static black hole (\ref{static-metric}) are located at the roots of 
$P(a)=0$. For $\Lambda>0$ and $C\geq 0$ there is only one real root $a^2=\xi_-^2R^2$, which in the 
$(y,t)$ coordinates translates to the curves
\be
\left(\frac{\bar r_\pm}{R}\right)^2= \frac{\lambda R^2}{4} +\frac{\xi_-^2R^2}{2a_0^2(t)}
\pm\sqrt{\left(\frac{\lambda R^2}{4}+\frac{\xi_-^2R^2}{2a_0^2(t)}\right)^2-
\left(\frac{\lambda R^2}{4}\right)^2-\frac{CR^2}{4a_0^4(t)}}.
\ee 
Figure \ref{fig-horizons} shows plots of these horizons for all allowed values of the constants $k$ and $\lambda$. 
\begin{figure}[htb]
	\centering
		\scalebox{0.7}{\includegraphics{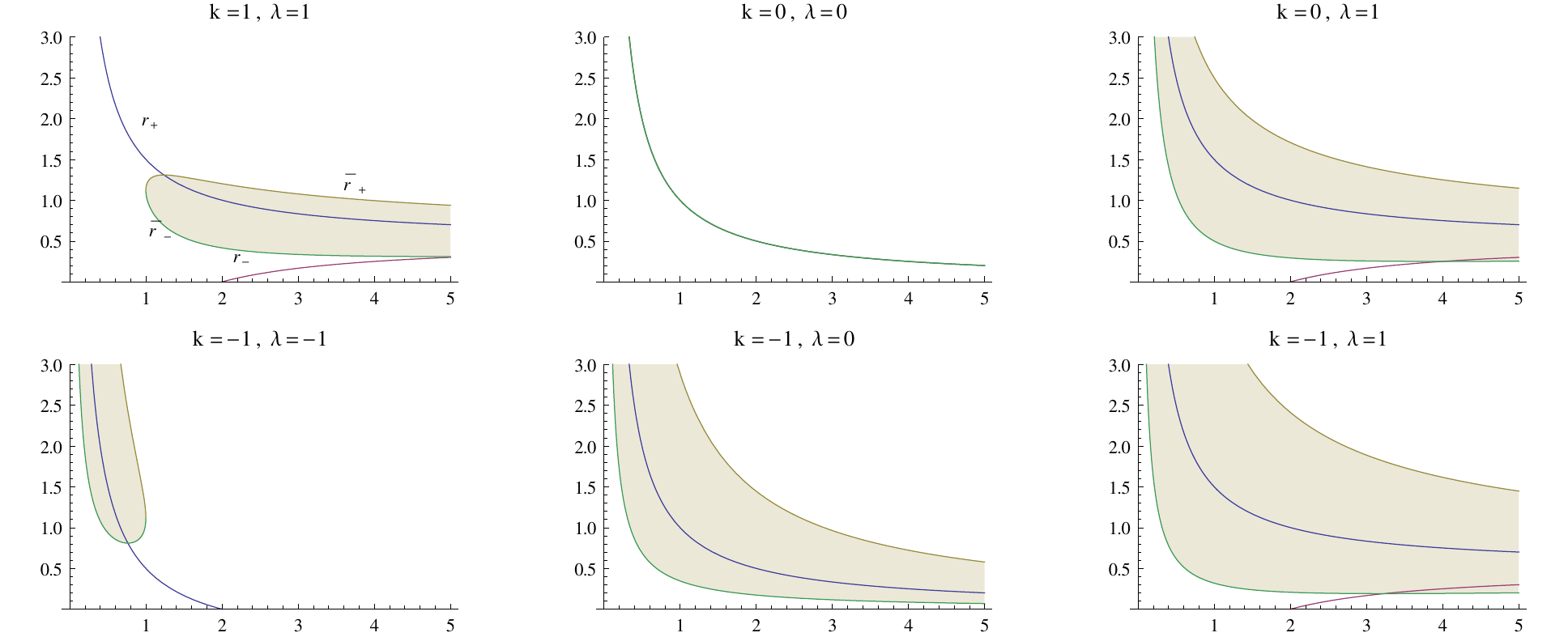}}
		\caption{Plots of the horizon curves $\left(\frac{r_\pm}{R}\right)^2$ and $\left(\frac{\bar r_\pm}{R}\right)^2$ versus $a_0^2(t)$ for $C=\Lambda=2$ and various values for $k$ and $\lambda$. 
		The shaded regions correspond to the regions where the diffeomorphism constructed above becomes singular, and hence	the $(a,\tau)$ coordinate system breaks down. For $k=\lambda=0$ the horizons 
		$r_+$ and $\bar r_\pm$ coincide, while for $k=\lambda=1$ and $k=\lambda=-1$ the regions $a_0<1$
		and $a_0>1$ respectively must be excised since they are non physical -- the solution ceases to be real. Due to the dilaton singularity at ${\bar r}_+$, the Wilson loop confines in the cosmological Liu-Tseytlin model \cite{EGM} in all cases except for $k=\lambda=0$ and $k=\lambda=1$. For $k=\lambda=0$, the dilaton singularity is also present, the string tension however vanishes since ${\bar r_+} = r_+$, and $n(t,r_+)=0$ as given in \eqref{tyequation} vanishes in this case. For $k=\lambda=1$ there is a  range of values $1 \leq a_0 \leq a_{0,\ast}$  for which $r_+ > {\bar r}_+$, in which the Wilson loop touches the $r_+$ horizon for large quark separation and therefore shows perimeter law behavior. The transition point is given by $a_{0,\ast}$ as in \eqref{a0star}. For later times $a_0 > a_{0\ast}$, the Wilson loop will touch the ${\bar r}_+$ horizon and shows area law behavior. In this case we hence find a Wilson loop confinement-deconfinement transition. For a cosmological slicing of AdS-Schwarzschild \cite{Tetradis} (i.e. constant dilaton backgrounds) the horizon structure is the same as for the cosmological Liu-Tseytlin model \cite{EGM}, in which case the Wilson loop show perimeter law behavor in all six cases. The situation is summarised in table~\ref{tab1}.}
		\label{fig-horizons}
\end{figure}

\section{Wilson Loops and the Quark-Antiquark Potential}\label{qqbar}

The \textbf{main result of \cite{EGM}} is the calculation of the static quark-antiquark potential from a fundamental string hanging from the boundary into the bulk space-time given by \eqref{y-co}, at any fixed time $t$ and as a function of proper distance $L$, measured in the FRW metric \eqref{FRW}. Following the standard procedure of calculating the embedding of a Nambu-Goto string, we found its tension
\beq\label{tension}
\tau_{q\bar q}(t) = \frac{n_s(r^*,t)}{2\pi\alpha'} = \frac{e^{\Phi(r,t)\over 2}}{2\pi\alpha'}\left| \frac{a(r,t) n(r,t)}{ a_0} \right|_{r^\ast}\,.
\eeq
The quark-antiquark potential scales with the distance $L$ if
$n_s(r,t)$ has a finite minimum at some distance $r=r^*$ outside the
horizon ${\bar r}_+$ at a given  $t$.
We now evaluate the Wilson loop tension from 
the general result  \eqref{tension} 
for the  dilaton solution given in \eqref{dilsol}, 
making use of additional information on the presence of an event
horizon with diverging dilaton \cite{LT}. By plotting \eqref{tension} numerically\footnote{A \textit{Mathematica} notebook can be found in the supplementary material to this ArXiv submission.}, we find that whenever ${\bar r_+}$ is the outermost horizon (i.e. the horizon with largest value of $r$) for a given ``time" $a_0$ (all cases except $k=\lambda=0$ and $k=\lambda=+1$ in fig.~\ref{fig-horizons}), $n_s(r)$ diverges to $+\infty$ due to the dilaton divergence at the horizon, and hence the Wilson loop confines.\footnote{As $r\rightarrow\infty$ the tension goes $\propto \frac{r^2}{R^2}$.} In the absence of the ${\bar r}_+$ event horizon the Wilson loop can  deconfine  whenever the naive horizon at $r_+$ can be reached, as $n(r_+,t)=0$, being the case for $k=\lambda=\pm 1$ in fig.~\ref{fig-horizons} for a interval $1 \leq a_0 \leq a_{0,\ast}$. If neither $r_+$ or ${\bar r}_+$ horizons are present, as for e.g. $k=\lambda=-1$, at large $a_0$ the Wilson loop would confine again as the tension is finite at $r=0$, where space ends in these coordinates. The cases $\lambda=k=0$ and $\lambda=k=+1$ are special:  In the former case the dilaton diverges at $r_+={\bar r}_+$, but the Wilson loop deconfines as this divergence is cancelled in  \eqref{tension} as $n(r,t)$ (c.f.~eq.~\eqref{tyequation}) approaches zero at the same location. The latter case, $\lambda=k=+1$, shows a \textit{Wilson loop confinement-deconfinement transition:} During the time interval $1 \leq a_0 \leq a_{0,\ast}$,
\beq\label{a0star}
a_{0,\ast}^2 = 1 - \sqrt{\frac{4 C}{R^2}} + \sqrt{1 + \frac{4 C}{R^2}}\,,
\eeq
 the Wilson loop will reach the naive horizon (i.e. the horizon given by $n(r_+,t)=0$), $r_+ \geq {\bar r}_+$, and exhibit perimeter law behaviour. For $a_0 > a_{0,\ast}$ on the other hand the Wilson loop will reach the event horizon ${\bar r}_+ > r_+$ and exhibit area law behaviour, due to the dilaton singularity. We summarize the situation in table~\ref{tab1}. We also note that the case   $k=\lambda=-1$ does not allow for deconfinement transitions, despite the presence of the $r_+$ horizon, since cosmological evolution requires $a_0^2 \leq 1$ such that the Wilson loop will not reach the naive $r_+$ horizon. In conclusion, except for the cases $k=\lambda=0$ and $k=\lambda=+1$, the Wilson loop is always confined during the cosmological evolution in the model of \cite{EGM}. For $k=\lambda=0$ the Wilson loop is always deconfined, and for $k=\lambda=+1$ we find a transition from a deconfined state at early times to a confined state at later times.}

\begin{table}\label{tab1}
	\centering
		\begin{tabular}{|c|c|c|c|}\hline
			& $\lambda=-1$ & $\lambda=0$ & $\lambda=+1$\\\hline
		$k=-1$ & \blue{C}, \red{D} & \blue{C}, \red{D} & \blue{C}, \red{D} \\\hline
	  $k=0$ & & \blue{D}, \red{D} & \blue{C}, \red{D} \\\hline 
	  $k=+1$ & & & \blue{D: $1\leq a_0 \leq a_{0,\ast}$, C: $a_0 > a_{0,\ast}$}, \red{D} \\\hline
		\end{tabular}
		\caption{(De)confinement (D/C) properties for cosmological \blue{Liu-Tseytlin} \cite{EGM} and \red{AdS/Schwarzschild} (i.e. constant dilaton) backgrounds \cite{Tetradis}. The stated properties hold for all values of $a_0$ attained during cosmological evolution unless stated otherwise. The value of $a_{0,\ast}$ is given in \eqref{a0star}.}
\end{table}

\section{Discussion}\label{disc}

We revisited and extended the holographic backgrounds of \cite{EGM} dual to $\mathcal{N}=4$ SYM on an FRW type cosmology and in the presence of a gluon condensate and instanton density, following the supersymmetric Ansatz of \cite{LT}. We found the most general solution of form \eqref{y-co} and introduced a (arbitrarily time-dependent) boundary cosmological constant $\lambda(t)$, which is equivalent to choosing an integration function in the bulk, the scale factor $a_0(t)$. We showed that the holographic stress-energy tensor is conserved for any $\lambda(t)$, clarifying conflicting claims between \cite{EGM} and e.g. \cite{AxioDilaton}. Furthermore, we found that these backgrounds are diffeomorphic to static black holes, whose event horizons become time-dependent in the cosmological frame, corresponding to a time-dependent observer viewing the static black hole event horizon. In the Liu-Tseytlin model \cite{LT} the dilaton diverges at the event horizon, covering the naive horizon $g_{tt}=0$ in most cases and in this way confining the Wilson loop. In two notable exceptions, $k=\lambda=0$ and $k=\lambda=+1$, the Wilson loop can reach the naive horizon $g_{tt}=0$ and deconfine. For $k=\lambda=+1$ this gives rise to a \textit{confinement-deconfinement} transition. On the other hand, in the absence of the gluon condensate, i.e. in a cosmological slicing of AdS/Schwarzschild with a constant dilaton, the Wilson loop is expected to break on the event horizon and to deconfine. Both cases are summarized in table~\ref{tab1}.

We conclude with several possibilities for future research: First, it would be interesting to construct other models with Wilson loop (de)confinement transitions (ideally at $k=0$, $\lambda>0$), and to apply the construction used here to other cases such as \cite{SS}. In such a model the free energy, another measure of (de)confinement, may also be calculated from holographic renormalization and linked to the Wilson loop behaviour. Finally, a further issue is the possible resolution of the dilaton singularity of \cite{LT}.


\begin{thebibliography}{[1]}

\bibitem{EGM}
  J.~Erdmenger, K.~Ghoroku and R.~Meyer,
  Phys.\ Rev.\  D {\bf 84} (2011) 026004.



\bibitem{Binetruy}
  P.~Binetruy, C.~Deffayet, U.~Ellwanger and D.~Langlois,
  Phys.\ Lett.\  B {\bf 477}, 285 (2000).

\bibitem{Tetradis}
  P.~S.~Apostolopoulos et.~al., 
  Phys.\ Rev.\ Lett.\  {\bf 102} (2009) 151301; 
  N.~Tetradis,
  JHEP {\bf 1003} (2010) 040.

\bibitem{CGKKM}
  C.~Charmousis, B.~Gouteraux, B.~S.~Kim, E.~Kiritsis and R.~Meyer,
  JHEP {\bf 1011} (2010) 151.

\bibitem{LT}
  H.~Liu and A.~A.~Tseytlin,
  Nucl.\ Phys.\  B {\bf 553} (1999) 231.

\bibitem{GGP} 
  G.~W.~Gibbons, M.~B.~Green and M.~J.~Perry,
  Phys.\ Lett.\  B {\bf 370}, 37 (1996).

\bibitem{Kehagias:1999vr}
  A.~Kehagias and E.~Kiritsis,
  JHEP {\bf 9911} (1999) 022.

\bibitem{KS2}
  A.~Kehagias and K.~Sfetsos,
  Phys.\ Lett.\  B {\bf 456}, 22 (1999).

\bibitem{Lang} 
  D.~Langlois,
  Phys.\ Rev.\  D {\bf 62}, 126012 (2000); 
  D.~Langlois and L.~Sorbo,
  Phys.\ Rev.\  D {\bf 68}, 084006 (2003).

\bibitem{AxioDilaton}
  I.~Papadimitriou,
  JHEP {\bf 1108} (2011) 119.


\bibitem{IPThermo}
    I.~Papadimitriou and K.~Skenderis,
  JHEP {\bf 0508} (2005) 004.

\bibitem{SS}
  T.~Sakai and S.~Sugimoto,   Prog.\ Theor.\ Phys.\  {\bf 113} (2005) 843; 
  Prog.\ Theor.\ Phys.\  {\bf 114} (2005) 1083.

\end{thebibliography}
\end{document}